\documentclass[twocolumn,5p,times,sort&compress]{elsarticle}

\usepackage{amssymb}
\usepackage{lineno}
\usepackage[usenames]{color}

\journal{Astroparticle Physics}

\begin{document}

\begin{frontmatter}

\title{A composition dependent energy scale and the determination of the cosmic ray primary mass in the ankle region}

\author[IAFE]{A. D. Supanitsky}
\ead{supanitsky@iafe.uba.ar}
\author[ITEDA]{A. Etchegoyen}
\author[ITEDA]{D. Melo}
\author[ITEDA]{F. Sanchez}
\address[IAFE]{Instituto de Astronom\'ia y F\'isica del Espacio (IAFE), CONICET-UBA, Buenos Aires, Argentina.}
\address[ITEDA]{Instituto de Tecnolog\'ias en Detecci\'on y Astropart\'iculas (CNEA, CONICET, UNSAM),
Centro At\'omico Constituyentes, San Martin, Buenos Aires, Argentina.}

\begin{abstract}

At present there are still several open questions about the origin of the ultra high energy cosmic rays.
However, great progress in this area has been made in recent years due to the data collected by the 
present generation of ground based detectors like the Pierre Auger Observatory and Telescope Array. In
particular, it is believed that the study of the composition of the cosmic rays as a function of energy 
can play a fundamental role for the understanding of the origin of the cosmic rays.   

The observatories belonging to this generation are composed of arrays of surface detectors and fluorescence
telescopes. The duty cycle of the fluorescence telescopes is $\sim 10\ \%$ in contrast with the $\sim 100\ \%$
of the surface detectors. Therefore, the energy calibration of the events observed by the surface detectors
is performed by using a calibration curve obtained from a set of high quality events observed in coincidence
by both types of detectors. The advantage of this method is that the reconstructed energy of the events
observed by the surface detectors becomes almost independent of simulations of the showers because just a
small part of the reconstructed energy (the missing energy), obtained from the fluorescence telescopes, 
comes from simulations. However, the calibration curve obtained in this way depends on the composition of
the cosmic rays, which can introduce biases in composition analyses when parameters with a strong dependence 
on primary energy are considered. In this work we develop an analytical method to study these effects. 
We consider AMIGA (Auger Muons and Infill for the Ground Array), the low energy extension of the Pierre Auger 
Observatory corresponding to the surface detectors, to illustrate the use of the method. In particular, we study 
the biases introduced by an energy calibration dependent on composition on the determination of the mean value 
of the number of muons, at a given distance to the showers axis, which is one of the parameters most sensitive 
to primary mass and has an almost linear dependence with primary energy.

\end{abstract}


\end{frontmatter}

%\linenumbers

\section{Introduction}

The cosmic ray energy spectrum extends over more than eleven orders of magnitude in energy (from below 
$\sim 10^9$ eV to above $\sim 10^{20}$ eV). It can be approximated by a broken power law with four spectral
features: the knee at a few $10^{15}$ eV \cite{Kulikov:58,Aglieta:04,Antoni:05,Apel:08,Amenomori:11}, the 
ankle at $\sim 4\times10^{18}$ eV \cite{Abu:01,Ave:01,Pravdin:03,HiRes:08,HiRes:09,Auger:08,Abu:13}, the 
cutoff or suppression at $\sim 3\times10^{19}$ eV \cite{HiRes:08,HiRes:09,Auger:08,Abu:13}, and a second 
knee at $\sim 10^{17}$ eV, recently reported by the KASCADE-Grande Collaboration \cite{KG:11}.

Several experimental techniques are used for the observation of the cosmic rays, depending on the energy
range under consideration. In particular, the direct observation of the primary particles is possible up 
to energies of the order of $\sim 10^{15}$ eV. For larger energies the study of cosmic rays is done by
observing the atmospheric air showers that they generate as a consequence of their interactions with air
molecules in the atmosphere. There are two classes of ground-based detectors, surface detectors and 
fluorescence telescopes. The surface detectors observe the lateral distribution of the showers by sampling
the secondary particles that reach the Earth's surface, whereas the fluorescence telescopes observe the 
fluorescence and Cherenkov photons generated, during the longitudinal development of the showers, as a
result of the interaction of the secondary charged particles with the air molecules
\cite{Baltrusaitis:85,Abu:00,AugerFD:10,Tokuno:12}.

Despite great experimental effort done in the last years the origin of the cosmic rays is still unknown. 
The observations used to study their origin comprise: the energy spectrum, the distribution of the arrival
directions, and the composition profile \cite{Stanev:11,Olinto:11}. 

Certainly, the detailed study of the composition as a function of energy is of great importance to unveil
the origin of the cosmic rays at all energies (see Ref.~\cite{Kampert2012a} for a review on composition). 
In particular, it is believed that the composition information is crucial to find the transition between 
the galactic and extragalactic components of the cosmic rays (see for instance Ref.~\cite{Medina-Tanco:07})
and to elucidate the origin of the suppression at the highest energies \cite{Kampert:13}. This feature of
the spectrum could originate as a result of the propagation of the cosmic rays in the intergalactic medium,
or by the end of the efficiency of the extragalactic sources to accelerate particles at the highest 
energies, or by a combination of both effects.

At the highest energies ($E\gtrsim10^{15}$ eV), the composition of the cosmic rays is studied by using
different observable parameters obtained from shower measurements which are very sensitive to the primary
mass. The parameters most sensitive to primary mass are the atmospheric depth at which the maximum
development of the showers is reached, which can be reconstructed from the fluorescence telescope data, 
and the muon content of the showers or a parameter closely related to it, which can be obtained from
dedicated muon detectors (see for instance Ref.~\cite{Supanitsky:09}).   

The fluorescence light emitted during the shower development is proportional to the deposited energy. A
fraction of these photons is detected by the fluorescence telescopes, making possible the reconstruction 
of the longitudinal profiles which yields an estimator of the energy of the primary particle. The energy
reconstructed in this way is largely independent of simulations, just a small correction is done, by using
simulations of the showers, which corresponds to the so-called invisible energy \cite{Song:00,Tueros:13}. 
In contrast, the energy calibration of surface detectors alone has to be done by using detailed simulations
of the showers and the detectors. The use of simulated showers introduces large systematic uncertainties
because the hadronic interactions at the energies of the cosmic rays are unknown. Then, the models used 
for shower simulation extrapolate low energy data taken from accelerator experiments by several orders of
magnitude in order to reach the energy of the cosmic rays. Note that, at present, the hadronic interaction
models are being updated in order to reproduce the Large Hadron Collider data, which reaches up to cosmic
ray energies of the order of $\sim 2 \times 10^{16}$ eV ($E_{cm}=7$ TeV)
\cite{Pierog:13,Ostapchenko:13,Pierog:13b,Csorgo:12,Khachatryan:10,Aamodt:10,Chojnacki:11,Chatrchyan:12}.   

The duty cycle of the fluorescence telescopes is $\sim 10\ \%$ and that of the surface detectors is 
$\sim 100\ \%$. Therefore, in order to have a large duty cycle and an energy calibration almost independent
of simulations, the present generation of cosmic ray observatories, the Pierre Auger Observatory in the
southern hemisphere and Telescope Array in the northern hemisphere, combine both techniques. The energy
scale of the surface detectors is obtained by using a parameter which is in general the interpolated signal
at a given distance to the shower axis. The calibration curve that relates this surface parameter with the
primary energy, reconstructed from the fluorescence telescope, is obtained experimentally from a subset of
high quality events observed in coincidence by both types of detectors \cite{Auger:08,Abu:13}.

Therefore, if the surface parameter used as an energy estimator depends on primary type, the energy scale 
depends on the composition of the cosmic rays. The use of this energy scale in composition analyses introduces 
biases that can be important when the parameters used to infer the primary mass have a strong dependence on 
primary energy (see Ref.~\cite{Ros:11} for a first study).

In this work we study the effects of using an energy scale dependent on composition in mass composition 
analyses. For that purpose we develop a dedicated analytical method. We consider the AMIGA project 
\cite{Etchegoyen:07} to illustrate in a simplified but realistic way the use of the method. The parameter 
sensitive to the primary mass is the number of muons at 600 m from the shower axis \cite{Supanitsky:08,Auger:04}, 
which depends almost linearly with primary energy. Therefore, composition analyses based on this parameter 
are supposed to be quite affected by the use of a composition dependent calibration curve.

It is worth mentioning that many surface parameters like $S_b$ \cite{Ros:11}, the risetime of the signals 
in surface detectors \cite{Ave:03,Auger:08RTRC}, the slope of the lateral distribution function 
\cite{Ave:03,Dova:04}, the curvature ratio of the shower front \cite{Auger:08RTRC,Rubtsov:13}, etc.~were 
proposed and sometimes used in composition analyses. The composition analyses that make use of these 
parameters together with a calibration curve dependent on composition are affected by the effect studied 
in this work for the case of the number of muons in the context of AMIGA. Each particular case involving 
different mass sensitive parameters and energy calibration methods has to be analysed in detail in order 
to estimate the importance of the biases introduced by this practice.

\section{Numerical approach}

\subsection{Simulations}

The simulations used in this work are the ones generated for the studies done in Ref.~\cite{Supanitsky:08}, 
they correspond to AMIGA. AMIGA will consist in a triangular grid of 750 m spacing composed by pairs of
detectors, a water-Cherenkov tank and a 30 m$^2$ muon counter buried underground. The energy region under 
consideration goes from $10^{17.6}$ eV up to $10^{18.5}$ eV. 

The atmospheric air showers used in Ref.~\cite{Supanitsky:08} to produce the simulated data were generated
by using AIRES \cite{aires} with QGSJET-II-03 \cite{qgsjetII} as the high energy hadronic interaction model.
The showers were simulated with fixed energies from $\log(E/\textrm{eV}) = 17.6$ to 
$\log(E/\textrm{eV}) = 18.5$ in steps of $\Delta \log(E/\textrm{eV}) = 0.1$. In this work just proton and 
iron primaries at $30^\circ$ zenith angle are considered. For each primary energy and primary type a set 
100 showers were generated. 

The muon counters are segmented in 192 scintillation strips. Part of the light generated by a charged 
particle that goes through a given strip is collected by a wavelength shifter fibre optic and transported 
to a multi-anode photomultiplier. The electronics has an independent channel for each strip which gives 
a digital one as output for each muon pulse. Note that more than one muon arriving in a time interval
corresponding to the width of a muon pulse is counted as one. That is called pile-up effect. 

As described in Ref.~\cite{Supanitsky:08}, a simplified simulation of the muon counters is performed.
Muon counters of 100 \% of efficiency buried underground at $2.5$ m depth, which corresponds to 
a muon energy threshold of $\sim 0.82$ GeV, are considered. The pile-up effect is also included in the 
simulations. 

From the data taken by the Cherenkov detectors the lateral distribution function of the 
signal, the impact point, and the arrival direction of the showers are reconstructed. While the muon 
lateral distribution function is reconstructed from the data taken by the muon detectors, following 
the method developed in Ref.~\cite{Supanitsky:08}. Note that the reconstruction method includes a 
correction for the pile-up effect.    

The parameter used as energy estimator, considered in this work, corresponds to the interpolated signal
collected by the water-Cherenkov detectors at 600 m from the shower axis, $S$, and the parameter sensitive
to primary mass corresponds to the interpolated number of muons at 600 m from the shower axis, $N_\mu$
\cite{AGASA:95}. Note that the signal $S$ is in units of VEM (vertical equivalent muon), which corresponds 
to the average signal deposited by a vertical muon that crosses the tank by its center \cite{Bertou:06}. 
Also note that $N_\mu$ is the total number of muons in an area of 30 m$^2$ for showers at $\theta=30^\circ$, 
therefore, it has no units. 

The AMIGA muon detectors were designed for zenith angles from $0^\circ$ to $45^\circ$. Therefore, we consider 
here showers of $30^\circ$ because it is the median of the zenith angle distribution. The parameter $N_\mu$ 
for showers of different zenith angles can be transformed to $30^\circ$ by using the corresponding muon 
attenuation curve \cite{Kascade:13}. 

\begin{figure}[th]
\centering
\includegraphics[width=9cm]{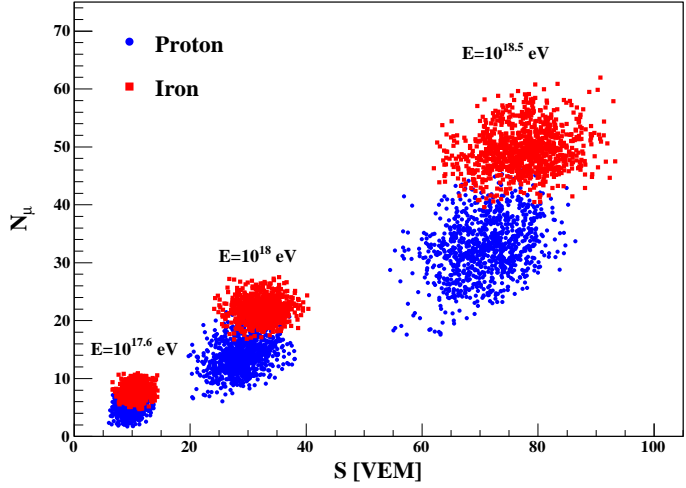}
\caption{$N_\mu$ versus $S$ for $E=10^{17.6}$, $10^{18}$, and $10^{18.5}$ eV corresponding to proton and 
iron primaries of $30^\circ$ of zenith angle. The high energy hadronic interaction model used is QGSJET-II-03. 
\label{NmuS}}
\end{figure}

Figure \ref{NmuS} shows $N_\mu$ versus $S$ for proton and iron primaries for three different values of
primary energy.  

The particle density at a given distance to the shower axis for the different types of particles of the showers 
presents shower to shower fluctuations. The corresponding distribution functions present asymmetric tails. 
However, the fluctuations introduced by the detectors and the effects of the reconstruction methods make a Gaussian 
function a good approximation of these distribution functions. Therefore, the combined distribution function of 
$N_\mu$ and $S$, given the primary energy and the primary type, can be approximated by a two-dimensional Gaussian 
function which is written as,
\begin{eqnarray}  
\label{Gauss2D}
P(N_\mu,S|E,A) \! \! \! &=& \! \! \! \frac{1}{2\pi\ \sigma[N_\mu]\ \sigma[S] \sqrt{1-\rho^2}} \nonumber \\
&& \times \exp \left[ -\frac{1}{2(1-\rho^2)} \left(\frac{(N_\mu-\langle N_\mu \rangle)^2}{\sigma^2[N_\mu]} \right. \right. \nonumber \\
&& +\frac{(S-\langle S \rangle)^2}{\sigma^2[S]} -2\rho
\nonumber \\ 
&& \left. \left. \times \frac{(N_\mu-\langle N_\mu \rangle)\ (S-\langle S \rangle)}{\sigma[N_\mu]\ \sigma[S]}% 
\right) \right],
\end{eqnarray}
where ($\langle N_\mu \rangle$, $\sigma[N_\mu]$) and ($\langle S \rangle$, $\sigma[S]$) are the mean value
and the standard deviation of $N_\mu$ and $S$, respectively. The correlation $\rho$ is given by,
\begin{equation}
\label{Cov}
\rho = \frac{\textrm{cov}(N_\mu,S)}{\sigma[N_\mu]\ \sigma[S]},
\end{equation}
where $\textrm{cov}(N_\mu,S)$ is the covariance between $N_\mu$ and $S$. Note that $\langle N_\mu \rangle$, 
$\langle S \rangle$, $\sigma[N_\mu]$, $\sigma[S]$, and $\rho$ are functions of primary energy ($E$) and
primary type ($A$).

\begin{figure}[h!]
\centering
\includegraphics[width=8.2cm]{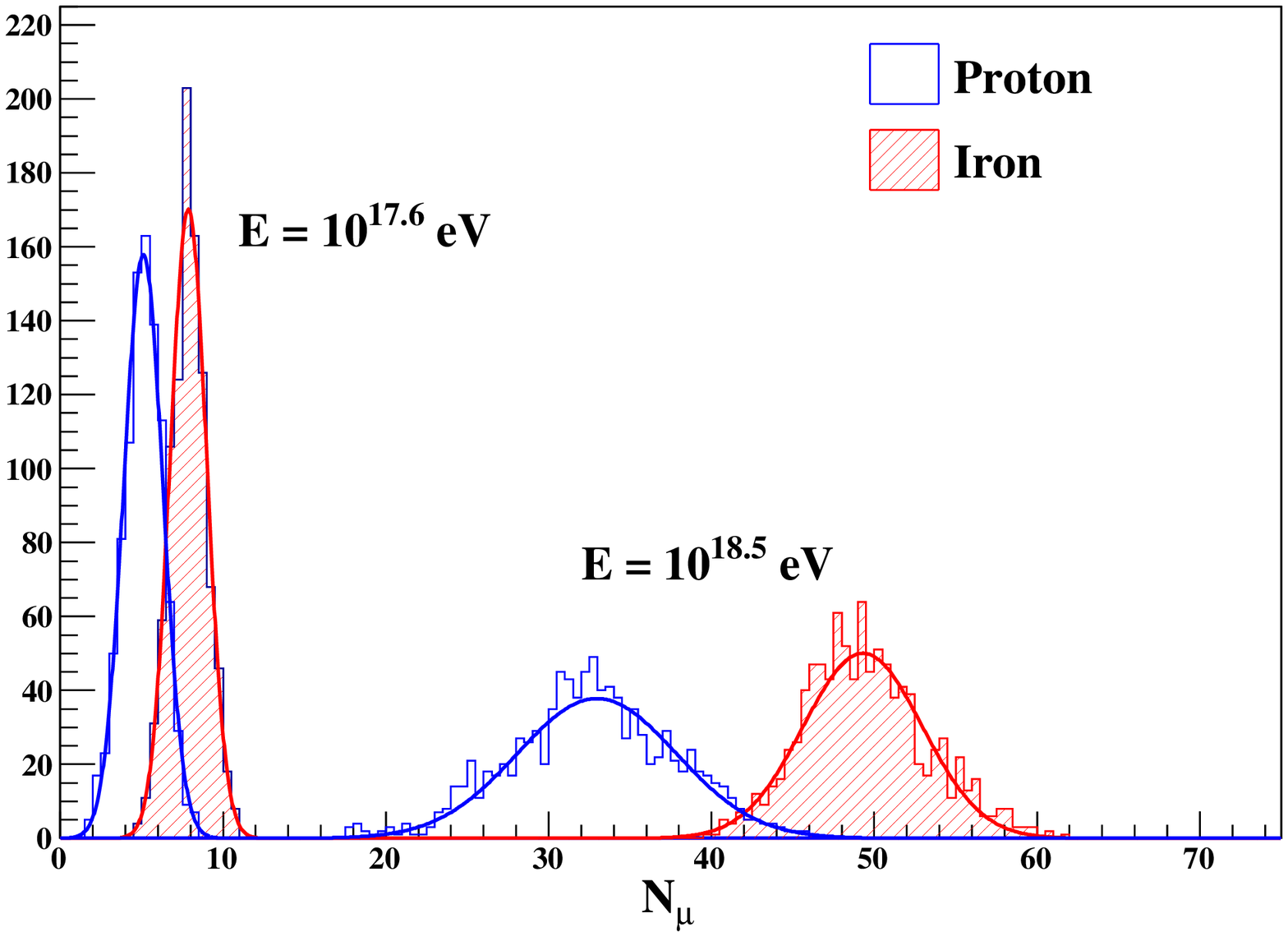}
\includegraphics[width=8.2cm]{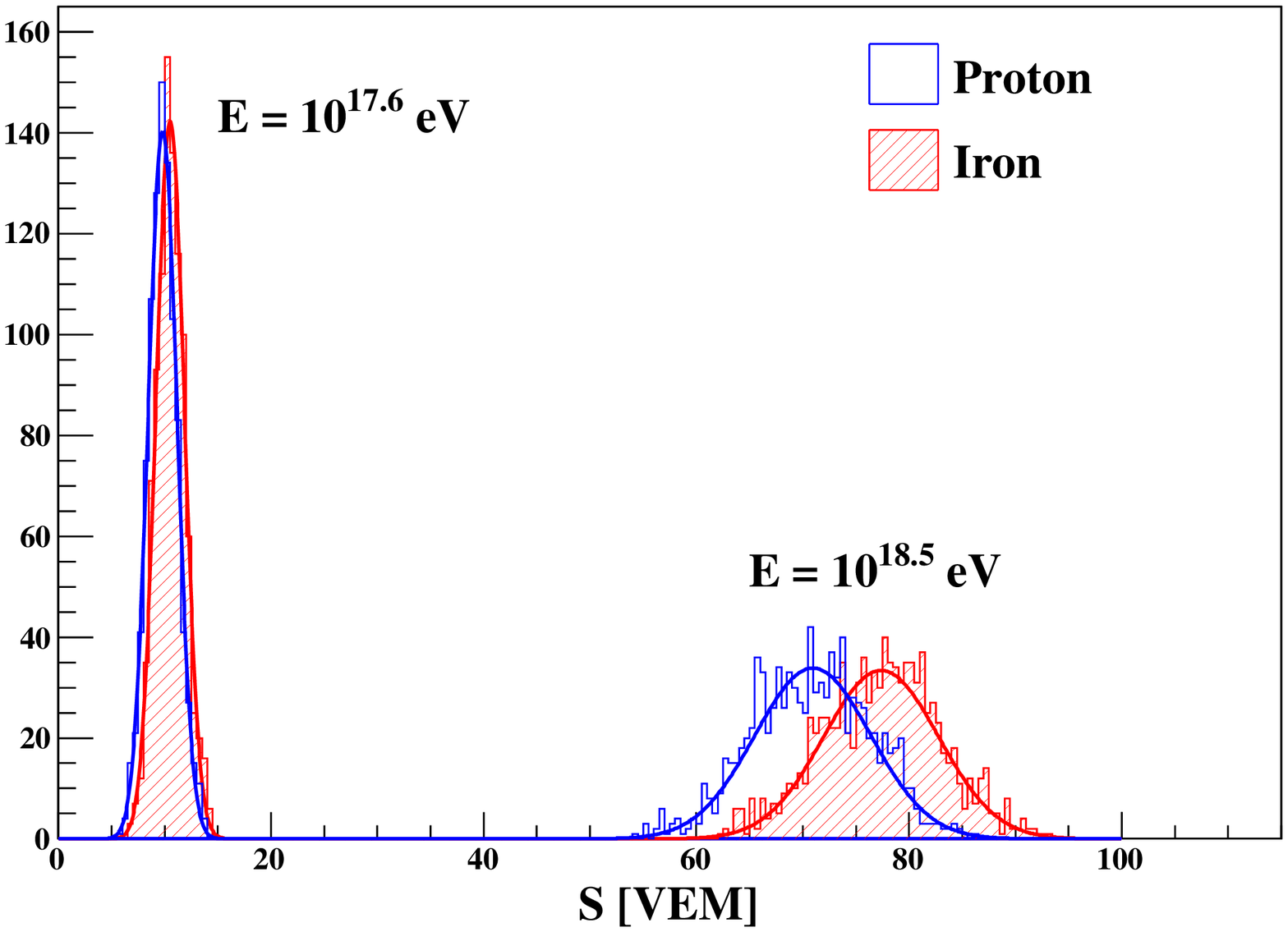}
\caption{One dimensional Gaussian fits to the proton and iron distributions of $N_\mu$ (top panel) and $S$ 
(bottom panel). The primary energies considered are: $E=10^{17.6}$ eV and $10^{18.5}$ eV. \label{Fits}}
\end{figure}

The parameters ($\langle N_\mu \rangle$, $\sigma[N_\mu]$) and ($\langle S \rangle$, $\sigma[S]$) for each 
primary energy and primary type are obtained by fitting the corresponding one dimensional distributions with 
one dimensional Gaussian functions. The correlation $\rho$ is obtained from the sample covariance and sample 
variances corresponding to each parameter (see Eq.~(\ref{Cov})). Figure \ref{Fits} shows the one dimensional 
Gaussian fits to the proton and iron distributions of $N_\mu$ (top panel) and $S$ (bottom panel) for two 
values of primary energy: $E=10^{17.6}$ eV and $10^{18.5}$ eV. It can be seen from the figure that the Gaussian 
fits are a good description of the distribution functions.

In order to obtain an analytical representation of $P(N_\mu,S|E,A)$ the logarithm of the mean value as a
function of the logarithm of primary energy and the standard deviation as a function of the logarithm of 
the primary energy, for each primary type and for each parameter ($N_\mu$ and $S$), are fitted. A linear
function is used in all cases except the one corresponding to $\sigma[N_\mu]$, for which a better fit is 
obtained with a quadratic function. Also, the logarithm of the correlation as a function of the logarithm 
of primary energy is fitted with a linear function for both types of primaries considered.

\begin{figure}[!h]
\centering
\includegraphics[width=8.2cm]{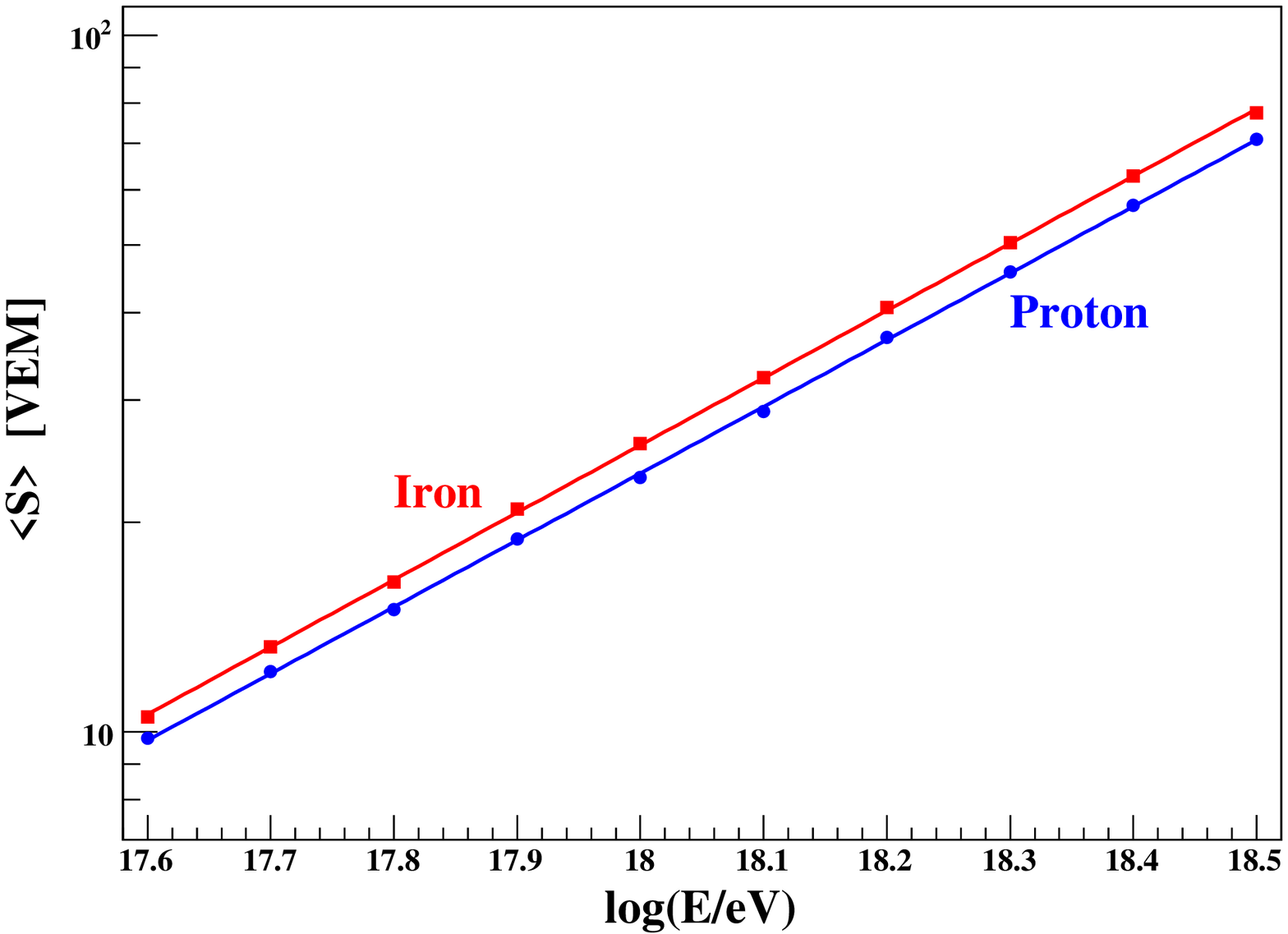}
\includegraphics[width=8.2cm]{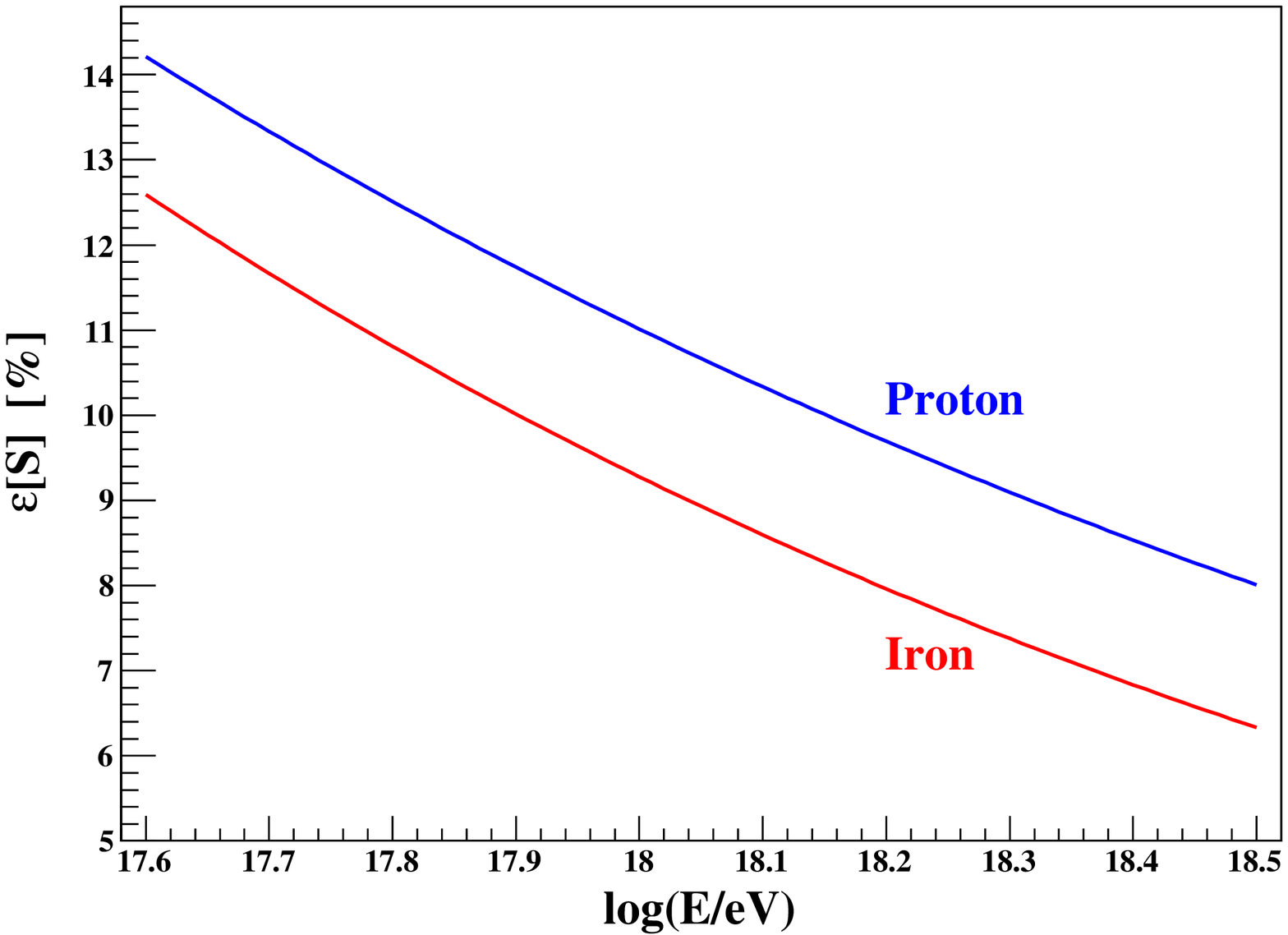}
\caption{Top panel: Mean value of $S$ as a function of the logarithm of primary energy for proton and 
iron primaries. Bottom panel: Relative error on the determination of $S$ as a function of the logarithm 
of primary energy for proton and iron primaries. \label{S}}
\end{figure}

The top panel of figure \ref{S} shows the mean value of $S$ as a function of the logarithm of primary 
energy for proton and iron primaries. Also shown are the fitted points (the error bars are included
but they are smaller than the markers). The difference between these two curves is nearly constant, it 
increases from $\sim 9\ \%$ to $\sim 11\ \%$ in the energy range considered. The bottom panel of figure
\ref{S} shows the relative error of $S$ which is given by, $\epsilon[S]=\sigma[S]/\langle S \rangle$. 
As expected, it decreases with primary energy for both primaries and it is smaller for iron nuclei in 
the whole energy range.   

There are experimental evidences about a deficit on the number of muons in simulated showers \cite{Farrar:13}.
As mentioned before, the hadronic interactions at the highest energies are unknown. As a consequence, 
models that extrapolate low energy data, obtained in accelerator experiments, to the energy of the cosmic 
rays are used. The signal in a water Cherenkov detector is the sum of the electromagnetic (due mainly to 
electrons, positrons, and gammas) and the muonic components. If the number of muons is larger than the 
one predicted by QGSJET-II-03, then $S$ should be more sensitive to the primary mass. Therefore, in order 
to consider more general cases, the difference between the mean value of $S$ for iron and proton primaries 
can be increased artificially. For that purpose let us introduce the parameter $\delta$ ($\delta \geq 0$)
such that,
\begin{eqnarray}  
\label{Delta}
\langle S \rangle (E,pr,\delta) \! \! \! &=& \! \! \! (1-\delta) \times \langle S \rangle (E,pr),  \nonumber \\
\langle S \rangle (E,fe,\delta) \! \! \! &=& \! \! \! (1+\delta) \times \langle S \rangle (E,fe).
\end{eqnarray}
The standard deviation of $S$ for both proton and iron nuclei is not modified. Note that $2 \times \delta$ corresponds 
to the fraction of the average $(\langle S \rangle (E,fe) + \langle S \rangle (E,pr))/2$ added to the difference between 
the mean value of $S$ for iron and proton primaries, as can be seen from the following equation,
\begin{equation}
\langle S \rangle (E,fe,\delta)-\langle S \rangle (E,pr,\delta) = \Delta_-\! \langle S \rangle(E) +
\delta \ \Delta_+\! \langle S \rangle(E),
\end{equation}
where $\Delta_\pm \langle S \rangle(E)=\langle S \rangle (E,fe) \pm \langle S \rangle (E,pr)$.

\begin{figure}[!h]
\centering
\includegraphics[width=8.2cm]{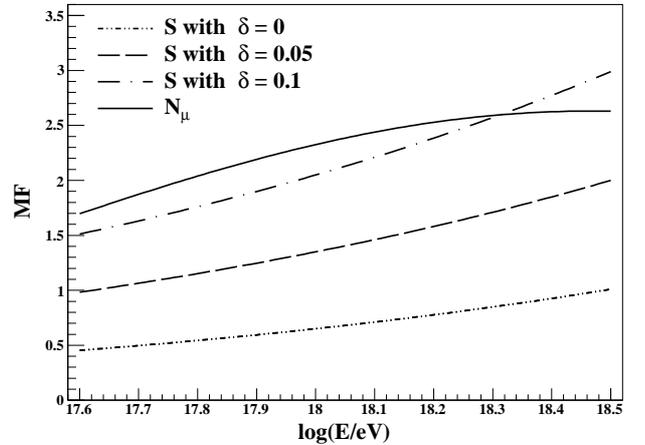}
\caption{Merit factor of $S$ as a function of the logarithm of primary energy for three different values 
of the parameter $\delta$: $0$, $0.05$, and $0.1$. The merit factor of $N_\mu$ is also shown. \label{MF}}
\end{figure}

The discrimination power of a given mass sensitive parameter, $q$, can be assessed by the commonly used
merit factor, which is defined as,
\begin{equation}
\textrm{MF}(q)=\frac{\langle q \rangle_{fe}-\langle q \rangle_{pr}}{\sqrt{\textrm{Var}[q]_{fe}+\textrm{Var}[q]_{pr}}},
\end{equation}
where $\textrm{Var}[q]_A$ is the variance of parameter $q$ for primary type $A$. Figure \ref{MF} shows the
merit factor of $S$ as a function of the logarithm of primary energy for three different values of 
$\delta$: $0$, $0.05$, and $0.1$. As expected the merit factor increases with primary energy (see figure 
\ref{S}). Also, from the figure it can be seen that for $\delta = 0$ the merit factor is smaller than one 
in the whole energy range under consideration, indicating that the discrimination power of $S$ is quite poor. 
For increasing values of $\delta$ the MF increases such that for $\delta = 0.1$ it is larger than $1.5$ in 
the whole energy range, reaching values close to $3$ for energies near $E=10^{18.5}$ eV. The figure also 
shows the merit factor of $N_\mu$ for comparison, which is of the order of the one corresponding to $S$ for
$\delta = 0.1$.

\subsection{Analysis of the bias}
\label{SecBias}

The mean value of the number of muons at a given distance to the shower axis is a parameter commonly 
used to infer the primary mass of the cosmic rays (see for instance \cite{CASAMIA,AGASA:05,Yakutsk:13}), 
due to its large sensitivity to the nature of the primary. In this section the effects on the determination 
of the mean value of $N_\mu$, introduced by the use of an energy scale dependent on composition are 
studied, considering a realistic physical situation. Also, a simplified case is analyzed in \ref{SC}
for which explicit analytical expressions of the most relevant quantities discussed here can be obtained.   

The calibration curve relates a parameter used as energy estimator (the signal $S$ in this case) with 
the reconstructed energy. Let us denote the calibration curve as $S_{cal}^{\mathcal{C}}(E_{rec})$ where
$\mathcal{C}$ indicates that this function depends on the composition of the cosmic rays. 

The flux of a given primary can be written as $J_A(E)=c_A(E)\ J(E)$, where $J(E)$ is the total flux and
$c_A(E)$ is the abundance corresponding to a primary of type $A$. Therefore, 
\begin{eqnarray}
\label{PAE}
P(A|E) \! \! \! &=& \! \! \! c_A(E), \\
\label{PEA}
P(E|A) \! \! \! &=& \! \! \! \frac{c_A(E)\ J(E)}{\int_0^\infty dE\ c_A(E)\ J(E)}, \\
\label{PE}
P(E) \! \! \! &=& \! \! \! \frac{J(E)}{\int_0^\infty dE\ J(E)},   
\end{eqnarray}
where $P(A|E)$ is the probability to find a nucleus of type $A$ given the true energy, $P(E|A)$ is the
energy distribution given the primary type, and $P(E)$ is the energy distribution of all species. Note 
that $\sum_A c_A(E) = 1$.

The combined distribution function of the number of muons and the reconstructed energy is given by,
\begin{equation}
\label{PNmuErec}
P(N_\mu,E_{rec}|E,A)=P(N_\mu,S_{cal}^{\mathcal{C}}(E_{rec})|E,A)\ \frac{\partial S_{cal}^{\mathcal{C}}}{\partial E_{rec}}(E_{rec}).
\end{equation}
Therefore, the distribution function of $N_\mu$ given the primary type and the reconstructed energy is
obtained from Eqs.~(\ref{PEA}) and (\ref{PNmuErec}),
\begin{eqnarray}
\label{Pnmu}
P(N_\mu | E_{rec}, A) \! \! \! &=& \! \! \! \frac{1}{M(E_{rec}, A)}\ \int^\infty_0 dE\ c_A(E)\ J(E) \nonumber \\
&& \times\ P(N_\mu,S_{cal}^{\mathcal{C}}(E_{rec})|E,A),
\end{eqnarray}
where, 
\begin{eqnarray}
\label{MPnmu}
M(E_{rec}, A) \! \! \! &=& \! \! \! \int_0^\infty dN_\mu\ \int^\infty_0 dE\ c_A(E)\ J(E) \nonumber \\
&& \times\ P(N_\mu,S_{cal}^{\mathcal{C}}(E_{rec})|E,A),
\end{eqnarray}
is the normalization of the distribution function. 

A power law energy spectrum is considered in all subsequent calculations. Therefore, the total flux can be 
written as,
\begin{equation}
J(E) = C\ E^{-\gamma}, 
\end{equation}   
where $C$ is a constant and $\gamma$ is the spectral index.

\subsubsection{Constant composition}

Let us first consider the simplified case in which there are just two nuclear species, proton and iron, 
and that the proton abundance $c_p$ is independent of primary energy. Assuming that the calibration curve 
is given by the mean value of the signal the following expression is obtained,   
\begin{equation}
\label{Scal}
S_{cal}^{\mathcal{C}}(E_{rec})=c_p\ \langle S \rangle (E_{rec},pr) + (1-c_p)\ \langle S \rangle (E_{rec},fe). 
\end{equation}
Note that the dependence on composition of the calibration curve is given explicitly. 

The signal $S$ corresponding to iron nuclei is larger than the one for protons (see figure \ref{S}).
Therefore, from Eq.~(\ref{Scal}) it can be seen that the reconstructed energy for iron nuclei is larger 
than the true one and for proton primaries is smaller. This bias in energy is translated into a bias in
composition analyses when a parameter with a strong dependence on primary energy, like the number of 
muons at ground, is considered.   

\begin{figure}[!h]
\centering
\includegraphics[width=8.2cm]{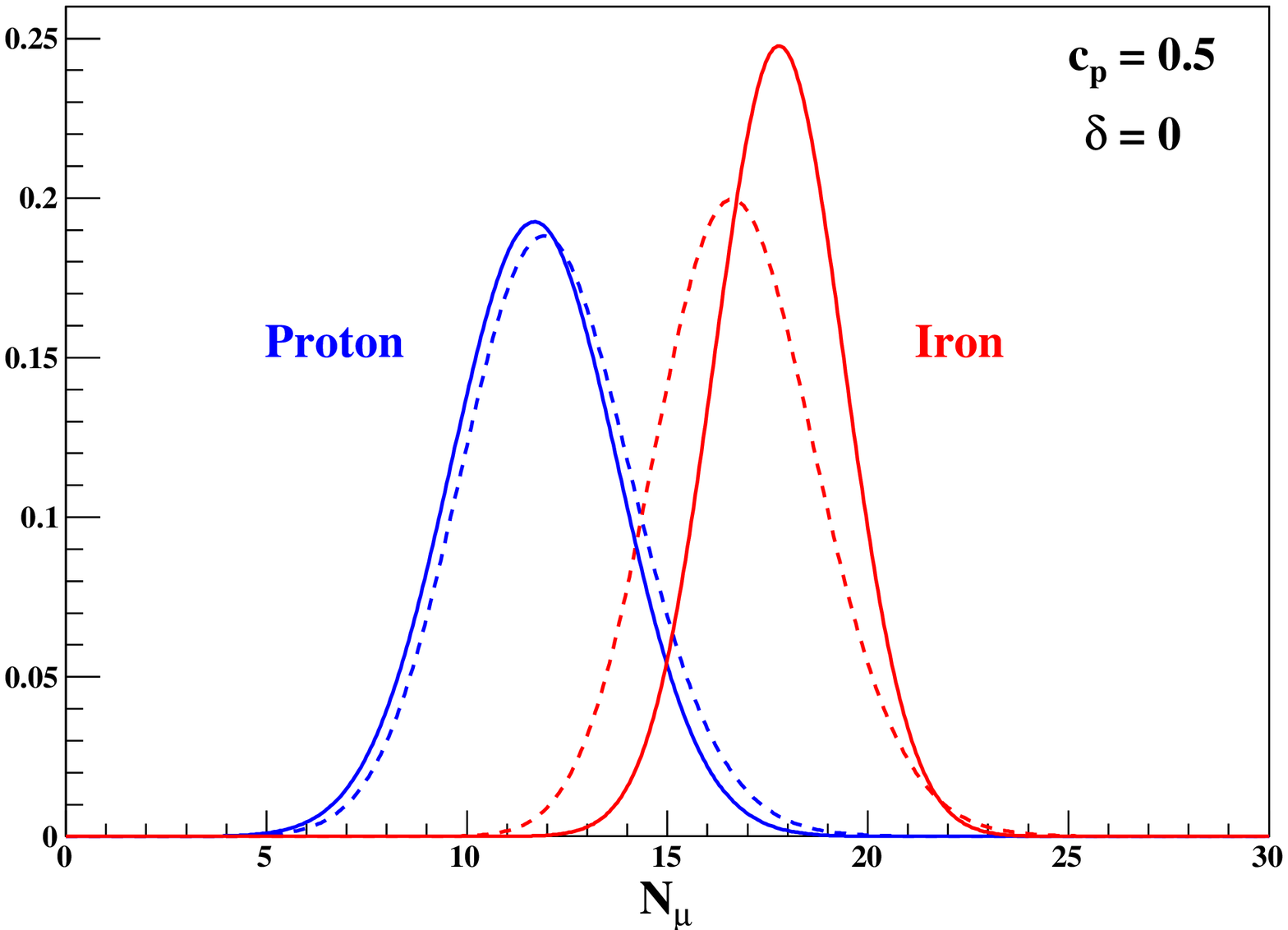}
\includegraphics[width=8.2cm]{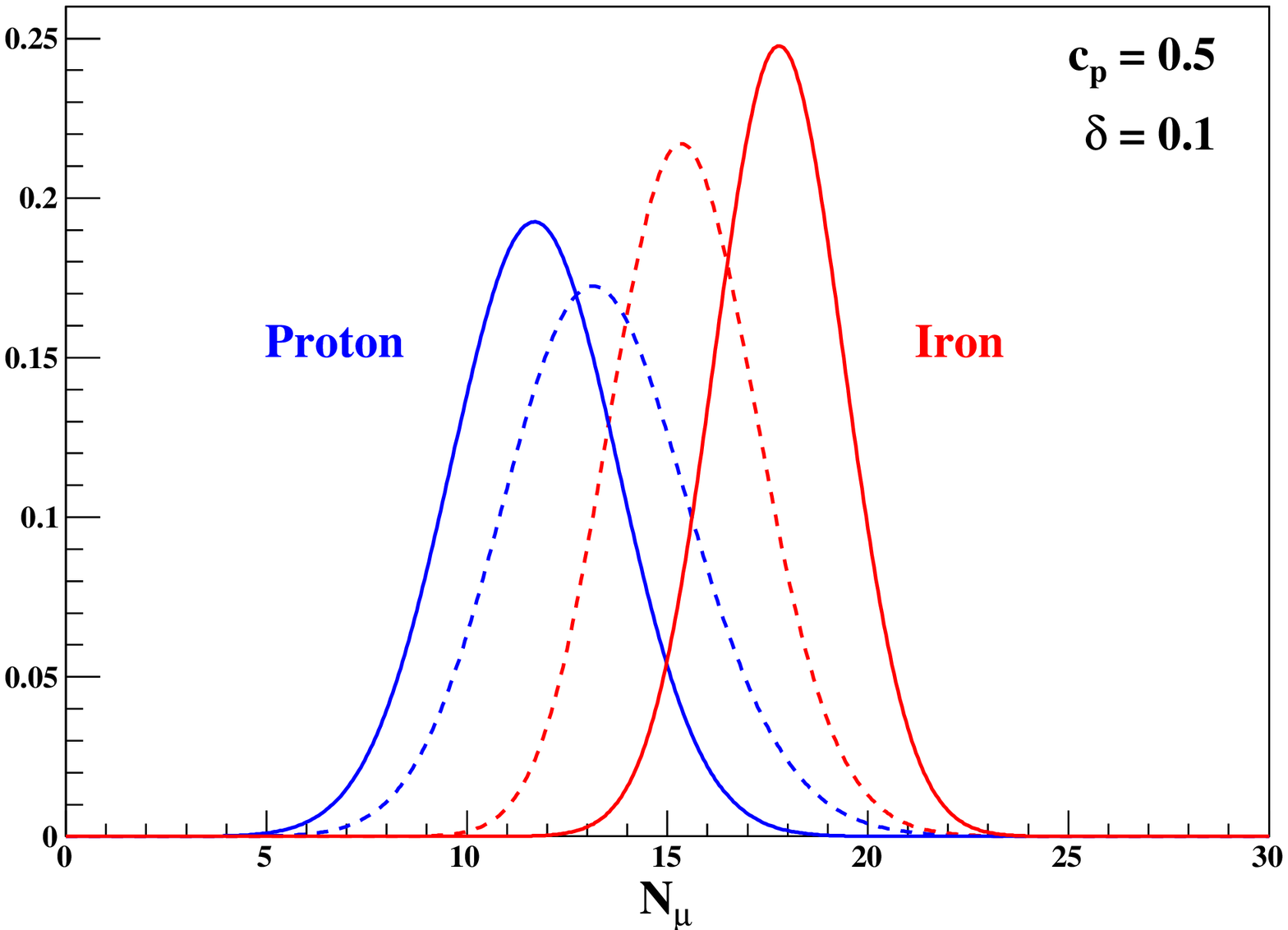}
\caption{Distribution functions of $N_{\mu}$ for proton and iron primaries of $E=E_{rec}=10^{18}$ eV. 
Top panel: $\delta=0$ and bottom panel: $\delta=0.1$. Solid lines correspond to the true energy and 
dashed line to the reconstructed energy. 
\label{NmuDist}}
\end{figure}

Figure \ref{NmuDist} shows the distribution functions of $N_\mu$ for proton and iron primaries for
$E=E_{rec}=10^{18}$ eV and for $\delta=0$ (top panel) and $\delta=0.1$ (bottom panel). The proton 
abundance considered is $c_p=0.5$ and the spectral index is $\gamma=3.27$, which corresponds to the 
experimental value obtained by The Pierre Auger Observatory in the energy range under consideration 
\cite{Ravignani:13}. It can be seen that the distribution functions of proton and iron primaries get 
closer. Note that the distribution function of iron nuclei is more affected than the one corresponding 
to proton primaries. This is because the calibration curve tends to move the distribution function of
protons to the right and the one corresponding to iron nuclei to the left, however, the energy spectrum
tends to move both distributions to the left.
Also, from the figure it can be seen that the modification of the distribution functions is more important
for increasing values of $\delta$, as expected.

The mean value of $N_\mu$ for a given primary type $A$, as a function of the reconstructed energy and for 
a proton abundance $c_A(E)$ can be calculated from Eq.~(\ref{Pnmu}),
\begin{eqnarray}
\label{MNmuG}
\langle N_\mu \rangle (E_{rec}, A) \! \! \! &=& \! \! \! \frac{1}{M(E_{rec}, A)}\ \int^\infty_0  dE \int^\infty_0 dN_\mu\ N_\mu \ c_A(E) \nonumber \\
&& \times\ J(E)\ P(N_\mu,S_{cal}^{\mathcal{C}}(E_{rec})|E,A),
\end{eqnarray}
which, by using Eq.~(\ref{Gauss2D}), takes the following form,
\begin{eqnarray}  
\label{MNmu}
\langle N_\mu \rangle (E_{rec},A) \! \! \! &=& \! \! \! \frac{1}{M(E_{rec},A)} \int_0^\infty dE\ c_A(E)\ J(E) \exp\Bigg[ \nonumber \\
&& -\frac{(S_{cal}^{\mathcal{C}}(E_{rec})-\langle S \rangle(E,A))^2}{2\ \sigma^2[S](E,A)} \Bigg] \nonumber \\
&& \times\ \Bigg( \langle N_\mu \rangle (E,A)+\rho(E,A)\ \frac{\sigma[N_\mu](E,A)}{\sigma[S](E,A)}  \nonumber \\
&& \times\ \big(S_{cal}^{\mathcal{C}}(E_{rec}) -\langle S \rangle(E,A) \big) \Bigg). 
\end{eqnarray}
Note that the correlation introduces a term which is directly added to the mean value of $N_\mu$ as a
function of the true energy. 

Therefore, from Eqs.~(\ref{PEA}), (\ref{PNmuErec}), and (\ref{MNmuG}) it can be demonstrated that the 
mean value of $N_\mu$ as a function of the reconstructed energy, corresponding to a mixture of nuclei, 
is given by,
\begin{equation}
\langle N_\mu \rangle (E_{rec}) = \sum_A \langle N_\mu \rangle (E_{rec},A)\ \omega_A(E_{rec}),
\label{Mnmu}
\end{equation}
where $\omega_A(E_{rec}) = M(E_{rec}, A)/ \sum_A M(E_{rec}, A)$. Note that for the ideal case in which the reconstructed
energy is equal to the true energy it easy to show that $\omega_A(E_{rec})=c_A(E_{rec})$.   

\begin{figure}[!h]
\centering
\includegraphics[width=8.2cm]{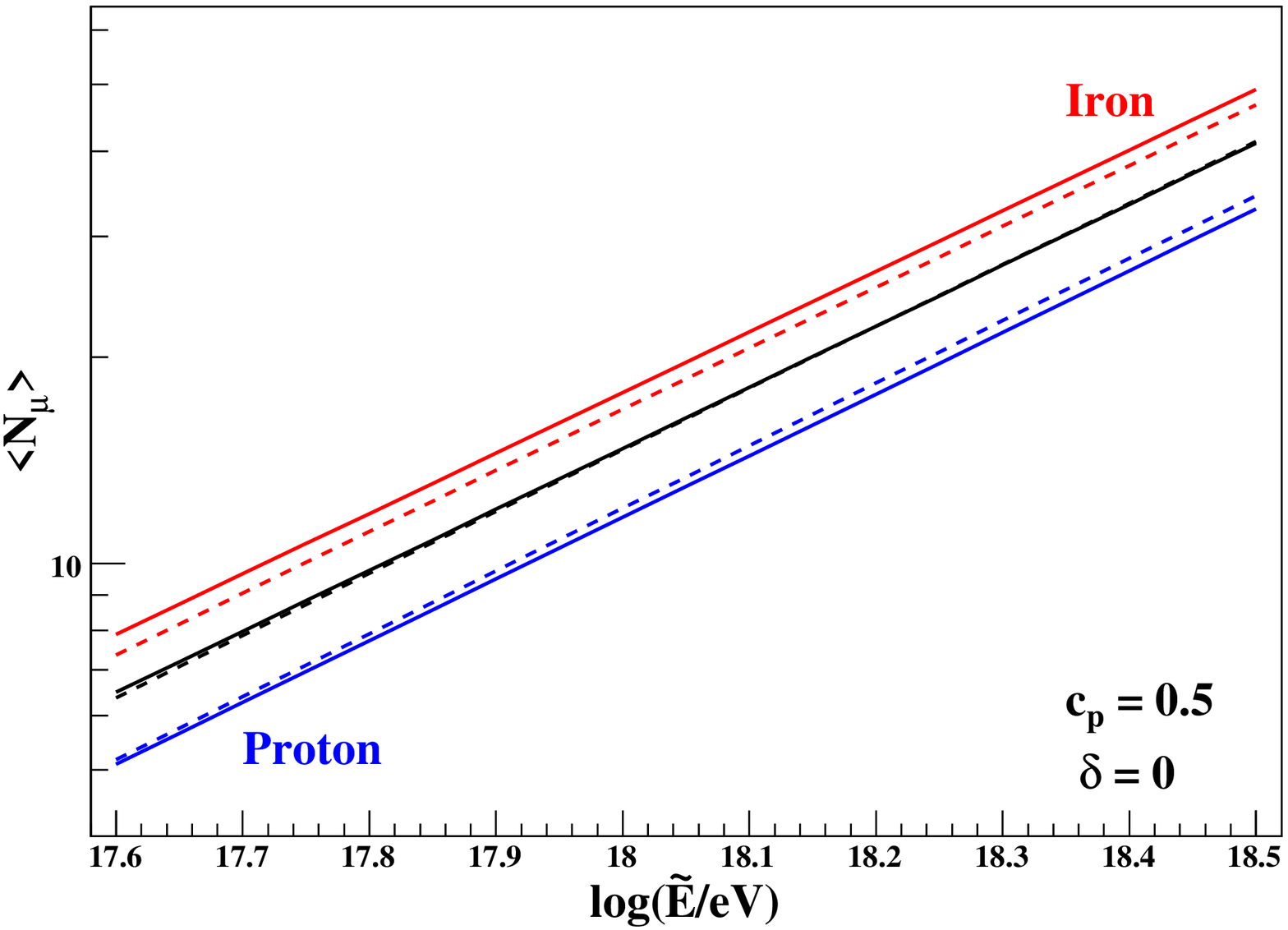}
\includegraphics[width=8.2cm]{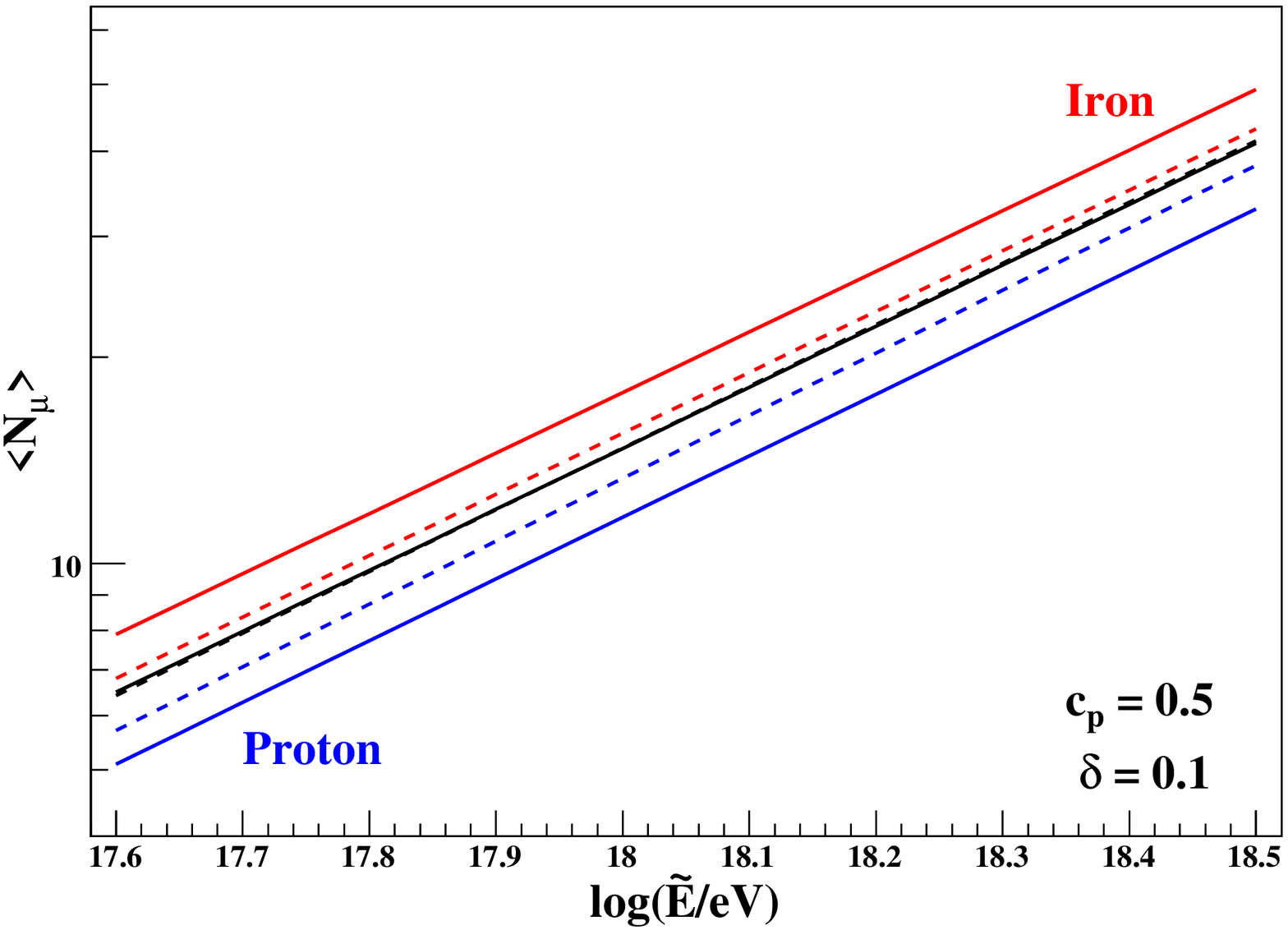}
\caption{Mean value of $N_\mu$ as a function of the logarithm of the energy for protons, iron nuclei, and 
a mixture of both primaries such that $c_p=0.5$. Top panel: $\delta=0$ and bottom panel: $\delta=0.1$. 
Solid lines correspond to the true energy and dashed lines to the reconstructed energy.
\label{MNmuDist}}
\end{figure}

Figure \ref{MNmuDist} shows the mean value of $N_\mu$ as a function of the logarithm of primary energy 
for protons, iron nuclei, and a mixture of both such that $c_p=0.5$ in the whole energy range. Also in 
this case, the solid lines correspond to the true energy ($E=\tilde{E}$) and the dashed lines correspond 
to the reconstructed energy ($E_{rec}=\tilde{E}$). 

From figure \ref{MNmuDist} it can be seen that the mean value of $N_\mu$ corresponding to each primary 
is affected by the dependence of the reconstructed energy on the proton abundance. In particular the mean
value corresponding to iron nuclei is underestimated and the one corresponding to protons is overestimated. 
This effect is quite large for the case of $\delta=0.1$. However when the mean value of the mixture is considered 
there is a cancellation that makes the difference between the true values and the ones
corresponding to the reconstructed energy small.

In order to quantify the difference between the true value of $\langle N_\mu \rangle$ and the one obtained
by using the reconstructed energy let us introduce the relative bias (see \ref{SC} for a simplified case 
in which the bias can be calculated explicitly), which is defined as,
\begin{equation}
\label{Rb}
R_b(\tilde{E}) = \frac{ \langle N_\mu \rangle (E_{rec}=\tilde{E}) }{ \langle N_\mu \rangle (E=\tilde{E}) }-1.
\end{equation}
Note that $R_b$ is positive for the case in which the true value of $\langle N_\mu \rangle$ is smaller 
than the one obtained by using the reconstructed energy.

\begin{figure}[!h]
\centering
\includegraphics[width=8.2cm]{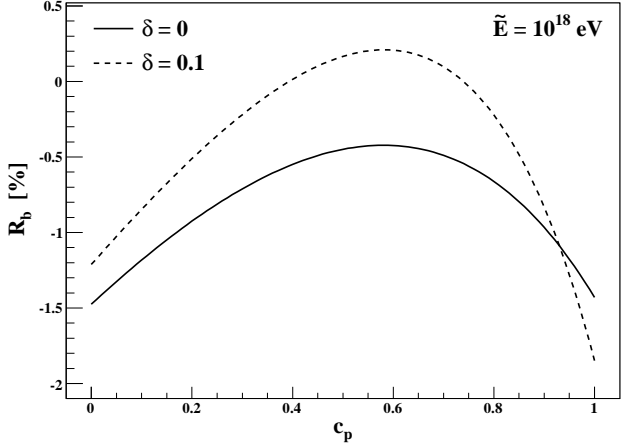}
\caption{Relative bias as a function of proton abundance for $\tilde{E}=10^{18}$ eV and for $\delta=0$ 
and $\delta=0.1$.
\label{BiasE18}}
\end{figure}

Figure \ref{BiasE18} shows the relative bias as a function of proton abundance for $\tilde{E}=10^{18}$ eV
and for $\delta=0$ and $\delta=0.1$. Both curves present a maximum between $c_p=0.5$ and $c_p=0.7$. The
relative bias corresponding to $\delta=0$ takes values between -1.5 \% and $-0.5$ \%, whereas the relative 
bias for $\delta=0.1$ takes values between -1.9 \% and 0.2 \%. Therefore, for $\delta=0.1$ the bias is
extended in a wider range than for $\delta=0$. In the extreme cases, $c_p=0$ and $c_p=1$, the bias comes
only from the convolution between the spectrum and the energy uncertainty because the composition is pure 
in both cases. From the figure it can be seen that for $\delta=0$ the absolute value of the relative bias 
is slightly larger for $c_p=0$. This is due to the larger muon content of iron showers, which is more 
important than the reduction of the absolute value of the bias for iron nuclei coming from the smaller 
energy uncertainty (see bottom panel of figure \ref{S}). For the $\delta=0.1$ case the absolute value of 
the relative bias for $c_p=0$ is smaller than the one corresponding to $c_p=1$. This is because for $\delta>0$
the relative error of $S$ increases for protons and decreases for iron nuclei ($\sigma[S]$ is kept constant).         

\begin{figure}[!h]
\centering
\includegraphics[width=8.2cm]{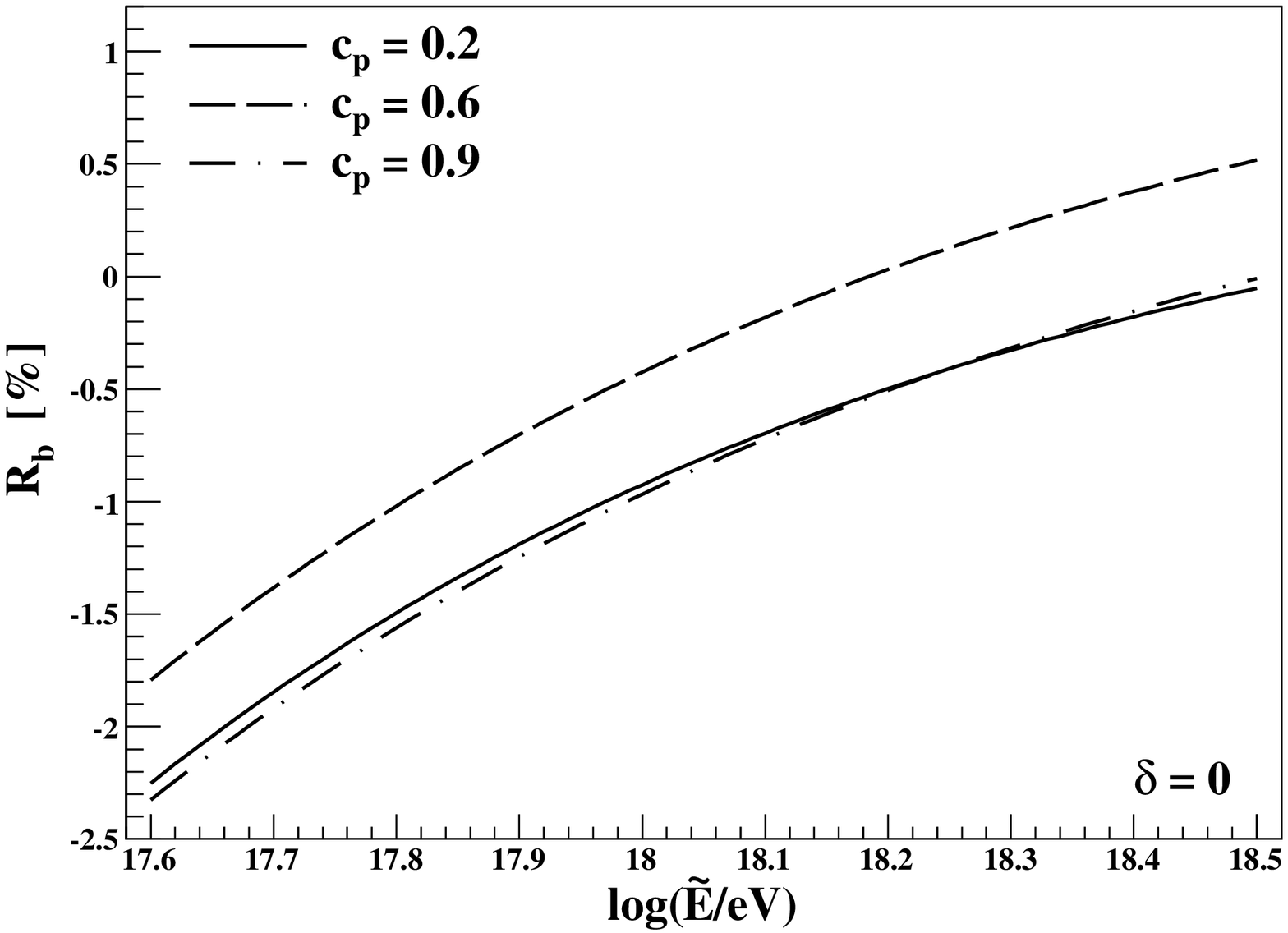}
\includegraphics[width=8.2cm]{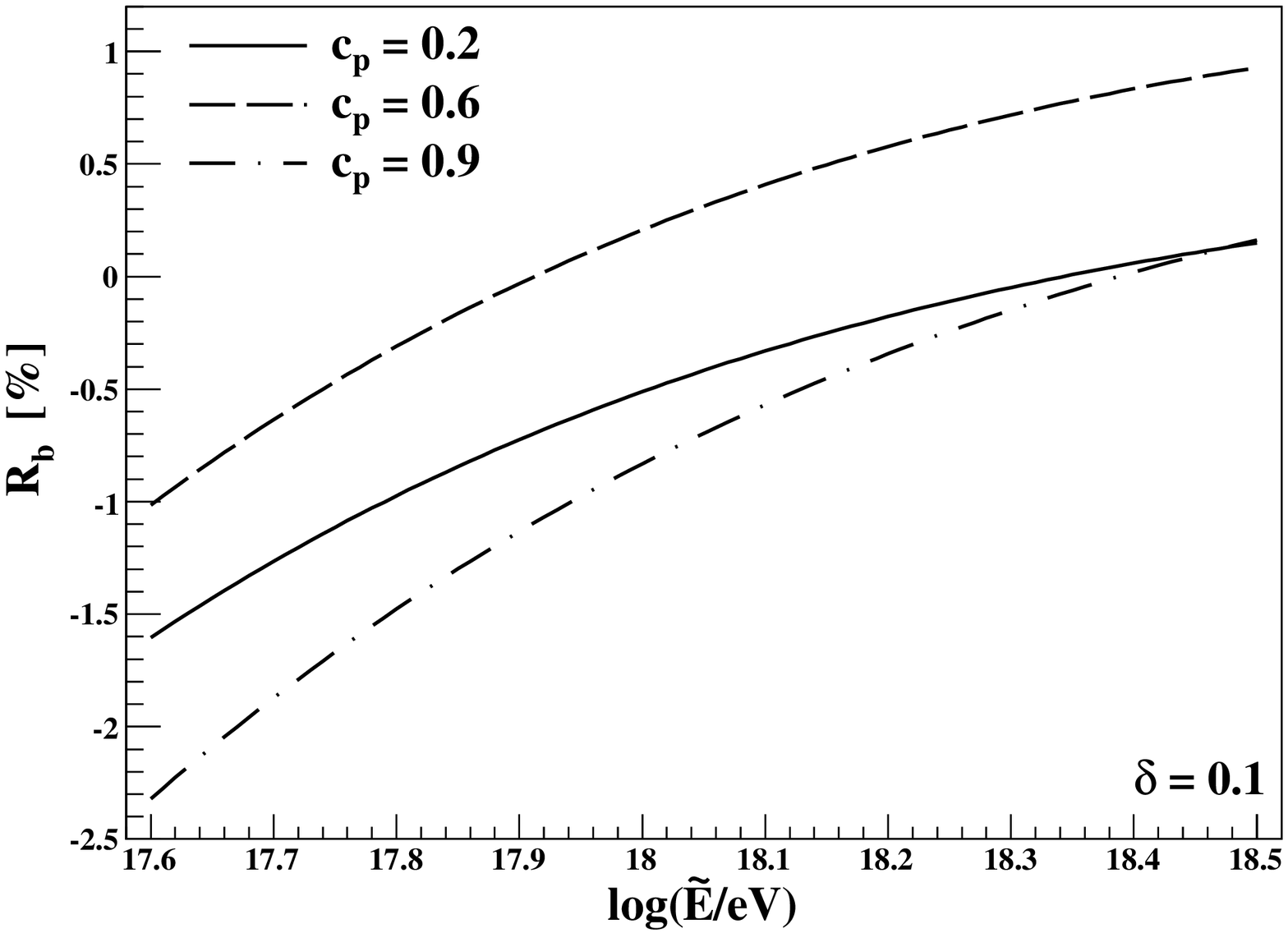}
\caption{Relative bias on the determination of $\langle N_\mu \rangle$ as a function of the logarithm of 
the energy. Top panel: $\delta=0$ and bottom panel: $\delta=0.1$. Three values of proton abundance are
considered, $c_p=0.2$, $0.6$, and $0.9$. 
\label{BiasMNmu}}
\end{figure}

Figure \ref{BiasMNmu} shows the relative bias on the determination of the mean value of $N_\mu$ as a
function of the logarithm of primary energy for $\delta = 0$ (top panel) and $\delta = 0.1$ (bottom panel)
and for $c_p = 0.2$, $0.6$, and $0.9$. The value of the spectral index used is the same as before, $\gamma=3.27$. 
For both values of $\delta$ and for all values of $c_p$ the absolute value of the relative
bias is smaller than $\sim 2.8\ \%$ in the energy range under consideration. This value corresponds to the
maximum of the module of the relative bias as a function of $c_p$ for $\tilde{E}=10^{17.6}$ eV.

\begin{figure}[!h]
\centering
\includegraphics[width=8.2cm]{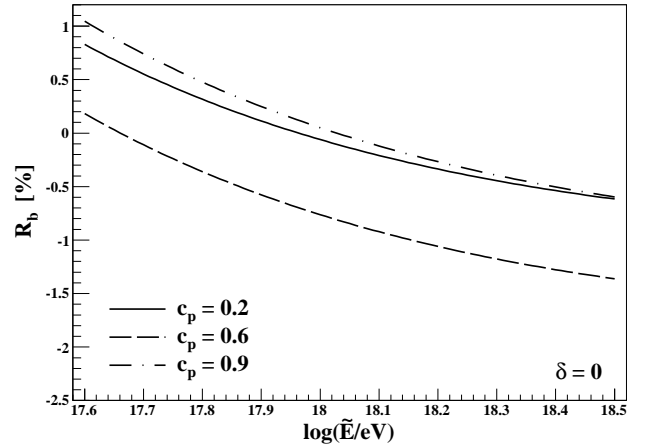}
\caption{Relative bias on the determination of $\langle N_\mu \rangle$ as a function of the logarithm of 
the energy for $\delta = 0$ and $\gamma=0$. Three values of proton abundance are considered, $c_p=0.2$,
$0.6$, and $0.9$.
\label{BiasMNmuGamma0}}
\end{figure}

As mentioned before, the bias on the determination of the mean value of $N_\mu$ depends on the spectral
shape of both primaries. Figure \ref{BiasMNmuGamma0} shows the relative bias as a function of the logarithm
of primary energy for $\gamma=0$ and for $\delta=0$. Comparing this figure with the top panel of figure
\ref{BiasMNmu} it can be seen that the relative bias has a quite different shape for different values of 
the spectral index $\gamma$. However, its absolute value is still quite small, less than $\sim 1.4\ \%$ in
this case.

For the case of $\gamma=0$ the decrease of the relative bias with primary energy is dominated by the 
slower increase of the mean value of $N_\mu$ as a function of the reconstructed energy, corresponding 
to iron primaries, compared with the one corresponding to the true energy. This is due to the fact that 
as the energy increases the merit factor of $S$ also increases (see figure \ref{MF}), then the bias 
coming from the dependence of the energy scale on $c_p$ is more important for increasing values of 
energy. The mean value of $N_\mu$ for iron primaries is more affected by this effect because the 
fluctuations of $N_\mu$ are smaller than the ones corresponding to protons. For the case of $\gamma=3.27$ 
the increase of the relative bias with primary energy is dominated by the faster increase of the mean 
value of $N_\mu$ as a function of the reconstructed energy, corresponding to proton primaries, compared 
with the one corresponding to the true energy. In this case the fast decrease of the energy spectrum with
primary energy tends to move the mean value of $N_\mu$ towards smaller values than the true ones but this
effect is gradually smaller as the energy increases because the fluctuations of $S$ decrease with primary
energy (see figure \ref{S}). As a consequence, the mean value of $N_\mu$ as a function of the reconstructed
energy corresponding to iron primaries is smaller than the one corresponding to $\gamma=0$ but the
difference with the true value as a function of primary energy is almost constant. For proton primaries 
the mean value of $N_\mu$ is also smaller compared with the one corresponding to $\gamma=0$ also due to 
the fast decrease of the energy spectrum. As for the case of iron nuclei this effect is less important 
for increasing energy making the mean value of $N_\mu$ to increase faster than the true one.  

\subsubsection{Varying composition}

Let us consider the case in which the composition profile depends on primary energy. For that purpose the
following shape for the proton abundance is assumed,
\begin{equation}
\label{cpE}
c_p(E) = \frac{1+\tanh(a\ \log(E/E_0))}{2}.
\end{equation}
It represents a transition from iron nuclei at low energies to protons at high energies. The transition 
is given at an energy $E_0$ and the speed at which this transition takes place is controlled by the
parameter $a$. The larger the values of $a$ the faster the transition from iron nuclei to protons. 

\begin{figure}[!h]
\centering
\includegraphics[width=8.2cm]{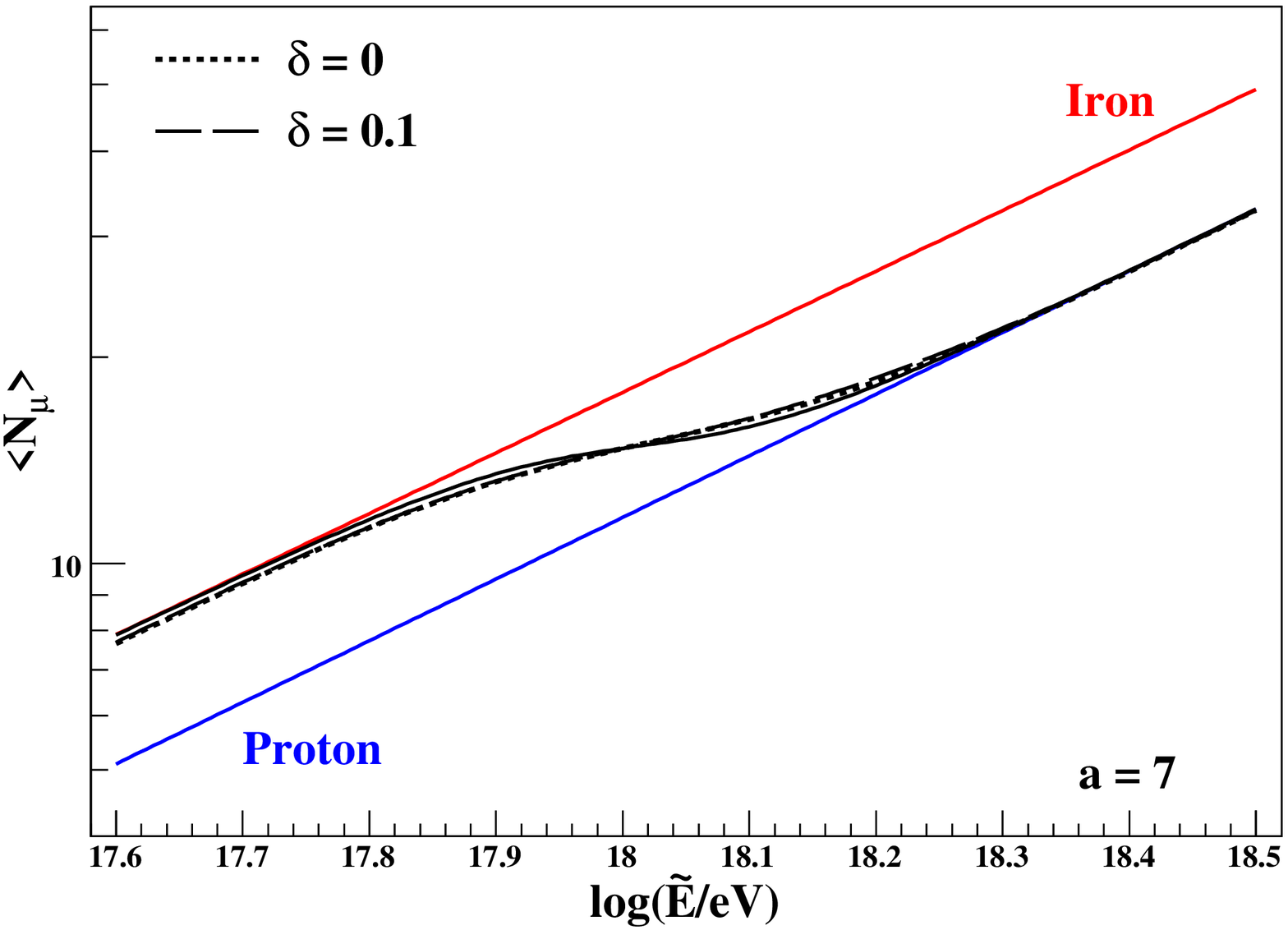}
\includegraphics[width=8.2cm]{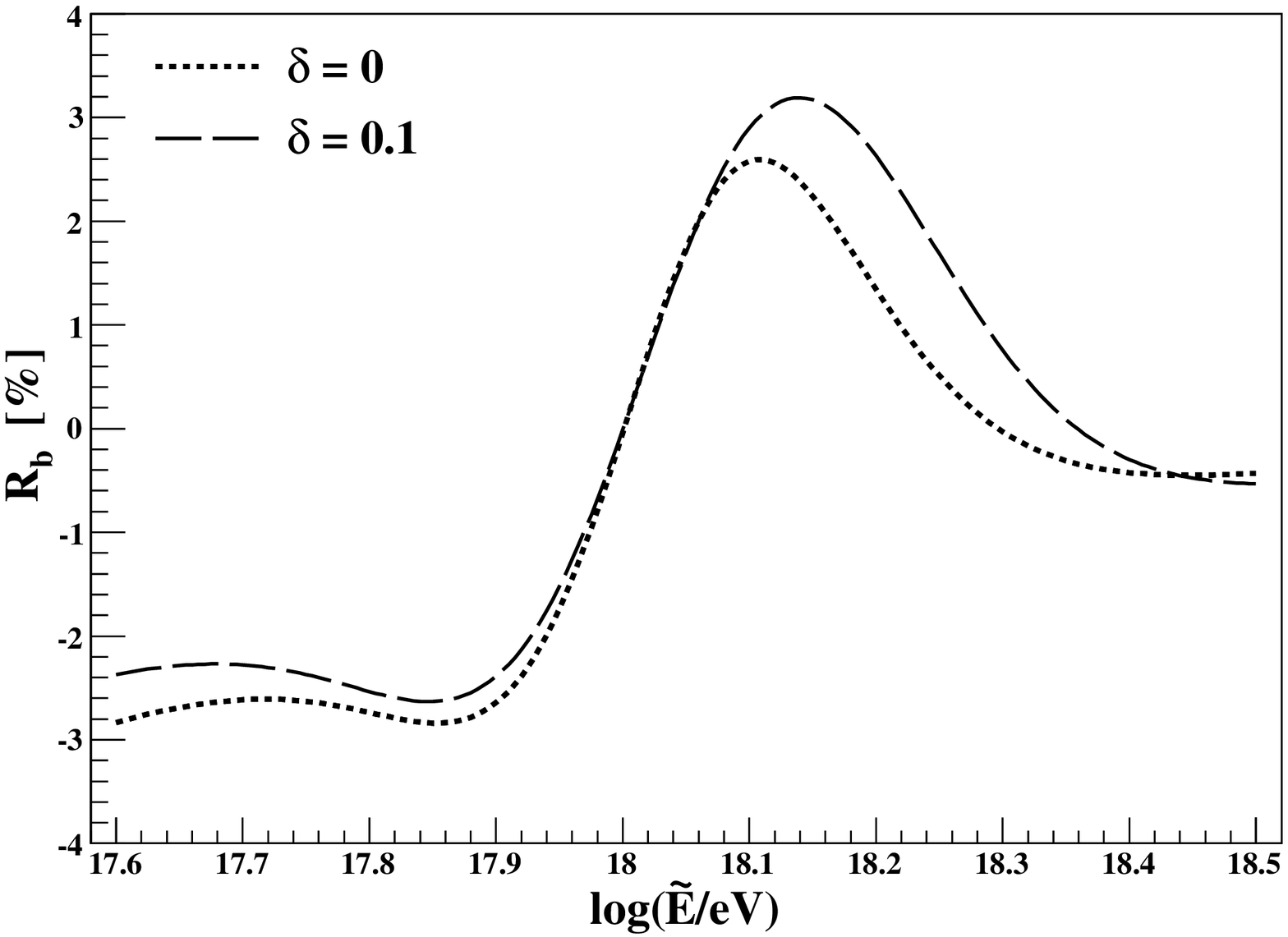}
\caption{Top panel: Mean value of $N_\mu$ as a function of the logarithm of primary energy. Bottom 
panel: Relative bias as a function of the logarithm of primary energy. The dotted and dashed lines
correspond to $\delta=0$ and $\delta=0.1$, respectively. The parameters corresponding to the composition
profile are: $E_0 = 10^{18}$ eV and $a = 7$ (see Eq.~(\ref{cpE})).
\label{BiasMNmuCpE07}}
\end{figure}

The top panel of figure \ref{BiasMNmuCpE07} shows the mean value of the number of muons as a function 
of the logarithm of primary energy for $E_0 = 10^{18}$ eV and $a = 7$. The calibration curve assumed for 
the calculation is given by Eq.~(\ref{Scal}) but in this case the proton abundance is a function of energy
(given by Eq.~(\ref{cpE})). When the reconstructed energy is considered (dashed and dotted lines) an energy
dependent bias appears. For both values of $\delta$ considered ($\delta=0$ and $\delta=0.1$) the transition
from iron nuclei to protons becomes slower than in the real composition profile. The bottom panel of figure
\ref{BiasMNmuCpE07} shows the corresponding relative bias as a function of energy for $\delta=0$ and
$\delta=0.1$. It can be seen that in both cases the relative bias takes values between $\sim -3\ \%$ and 
$\sim 3.2\ \%$. Note that for $\delta = 0.1$ the relative bias expands over a slightly larger region than for
$\delta = 0$ but the difference is small.      

\section{Conclusions}

In this work we have studied the importance of a composition dependent energy scale on composition analyses. 
The method pursued dwells on a combined distribution function of the mass and energy estimator parameters 
which allows to analytically perform all further analyses and their physical interpretation. In this paper 
we have shown the strength of this approach which might be applied to different experimental scenarios with 
appropriate distribution function. This approach allows a clear insight in the impact of different mass 
composition mixing, which is of paramount importance to understand the cosmic ray spectrum, in particular 
in composition changing regions.   

We have applied the method developed to AMIGA in order to exemplify these effects in a realistic experimental 
context. We have taken the number of muons and the signal in the water-Cherenkov detectors, both evaluated at 
600 m from the shower axis, as the mass sensitive and energy estimator parameters, respectively. We have found 
that the distribution functions of the number of muons for proton and iron primaries can be modified when an 
energy calibration dependent on composition is used to reconstruct the energy of the events. However, the 
relative bias on the determination of the mean value of the number of muons is quite small, of the order of 
a few \% in the whole energy range under consideration. This is true even for energy estimators with merit 
factors of the order of the one corresponding to the number of muons. We have obtained the same upper limit 
on the relative bias for the two physical situations that we have considered: a constant proton abundance as 
a function of primary energy and a smooth transition from iron to proton primaries at $E=10^{18}$ eV.

It is worth mentioning that the impact of the use of an energy scale dependent on composition in composition
analyses has to be analyzed in detail in each particular case. The reason for that is that the effects 
introduced by this practice depend on the parameter sensitive to the primary mass under consideration, on the 
type of detectors used to observe the air showers, and on the methods used to reconstruct the shower parameters.

\section{Acknowledgements}

The authors have greatly benefited from discussions with several colleagues from the Pierre Auger
Collaboration, of which they are members. We specially thank Roger Clay for the review of the manuscript.
The authors are members of the Carrera del Investigador Cient\'ifico of CONICET, Argentina. This work 
is supported by CONICET PIP 114-201101-00360 and ANPCyT PICT-2011-2223, Argentina.

\appendix

\section{A simplified case}
\label{SC}

In the simplified physical situation treated here it is assumed that the fluctuations of the parameter 
used to reconstruct the primary energy are negligible, i.e.~$\sigma[S](E,A)=0$ for $A=pr$ and $fe$. 
Considering that the energy calibration is given by the mean value of the signal, it is possible to 
obtain the true energy $E$ of each primary as a function of the reconstructed energy from the following
expression,
\begin{equation}
\label{ScalSC}
\langle S \rangle(E,A)=c_p \langle S \rangle(E_{rec},pr) + (1-c_p) \langle S \rangle(E_{rec},fe).
\end{equation}
Here a binary mixture of protons and iron nuclei and a constant proton abundance are assumed.

The mean value of the number of muons is obtained, in this case, following a similar procedure to the 
one described in section \ref{SecBias},
\begin{eqnarray}
\langle N_\mu \rangle(E_{rec}) \! \! \! &=& \! \! \! \bigg( c_p\ \langle N_\mu \rangle(E(E_{rec},pr),pr)\ J(E(E_{rec},pr))  \nonumber \\
&& \times\ \frac{\partial E}{\partial E_{rec}}(E_{rec},pr)+(1-c_p) \nonumber \\ 
&& \times\ \langle N_\mu \rangle(E(E_{rec},fe),fe)\ J(E(E_{rec},fe)) \nonumber \\  
&& \times\ \frac{\partial E}{\partial E_{rec}}(E_{rec},fe) \bigg) \times \bigg(c_p\ J(E(E_{rec},pr)) \nonumber \\
&& \times\ \frac{\partial E}{\partial E_{rec}}(E_{rec},pr)+(1-c_p)\ J(E(E_{rec},fe)) \nonumber \\
\label{MNmuSC}
&& \times\ \frac{\partial E}{\partial E_{rec}}(E_{rec},fe) \bigg)^{-1}. 
\end{eqnarray}

The mean values of parameters $N_\mu$ and $S$ have an almost linear dependence on primary energy. Then, 
in order to further simplify the calculation let us assume an exact linear dependence on primary energy 
of the mean value of both parameters,
\begin{eqnarray}
\label{NmuSC}
\langle N_\mu \rangle(E,A) \! \! \! &=& \! \! \! N_{\mu,0}^A \left(\frac{E}{E_0}\right), \\
\label{SSC}
\langle S \rangle(E,A) \! \! \! &=& \! \! \! S_{0}^A \left(\frac{E}{E_0}\right),
\end{eqnarray}
where $E_0$ is a reference energy. Therefore, the mean value of $N_\mu$ as a function of the true 
energy is given by,
\begin{equation}
\label{NmuSCT}
\langle N_\mu \rangle(E)=\left(c_p\ N_{\mu,0}^{pr} + (1-c_p)\ N_{\mu,0}^{fe}\right) \left(\frac{E}{E_0}\right).  
\end{equation} 

By using Eqs.~(\ref{ScalSC}, \ref{MNmuSC},\ref{NmuSC},\ref{SSC}) and (\ref{NmuSCT}) the relative bias 
on the mean value of the number of muons as a function of the reconstructed energy can be written as,
\begin{equation}
\label{RbSC}
R_b = \frac{(R_\mu-R_S)\ (R_S^{\gamma-2}-1)\ c_p\ (1-c_p)}{\left(c_p+R_\mu\ (1-c_p)\right)\ \left(c_p+R_S^{\gamma-1}\ (1-c_p)\right)},
\end{equation}
where,
\begin{eqnarray}
R_\mu \! \! \! &=& \! \! \! \frac{N_{\mu,0}^{fe}}{N_{\mu,0}^{pr}}, \\
R_S \! \! \! &=& \! \! \! \frac{S_{0}^{fe}}{S_{0}^{pr}}.
\end{eqnarray}
Here a power law energy spectrum of the form $J(E)=C\ E^{-\gamma}$ is assumed.

\begin{figure}[th]
\centering
\includegraphics[width=8.2cm]{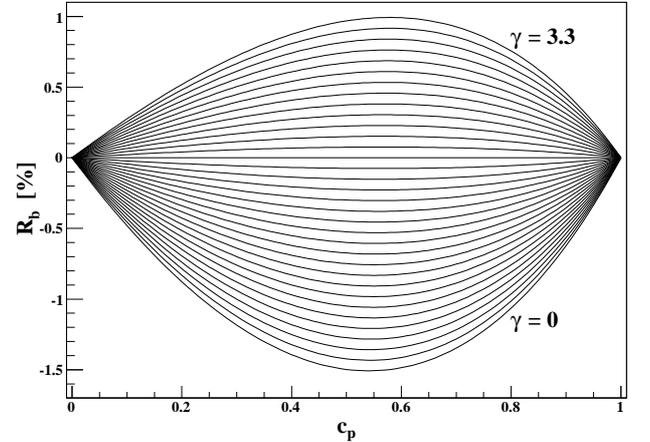}
\caption{Relative bias as a function of proton abundance for different values of the spectral index 
starting from $\gamma=0$ up to $\gamma=3.3$ in steps of $\Delta \gamma = 0.1$. \label{BiasSC}}
\end{figure}

From Eq.~(\ref{RbSC}) it can be seen that the relative bias is independent on primary energy. Also, 
when the composition is pure, the cosmic rays are only protons or iron nuclei ($c_p=1$ or $c_p=0$), 
the relative bias is zero, as expected. Moreover, the bias also disappears when $R_S=1$, $R_\mu = R_S$, 
or $\gamma=2$. Figure \ref{BiasSC} shows the relative bias as a function of the proton abundance for
different values of $\gamma$, starting from $\gamma=0$ up to $\gamma=3.3$ in steps of 
$\Delta \gamma = 0.1$. The values $R_S = 1.1$ and $R_\mu=1.5$ are used to make the plot, which correspond 
to the ratios between the mean values of each parameter ($N_\mu$ and $S$) for proton and iron at 
$E=10^{18}$ eV.

From figure \ref{BiasSC} it can be seen that the relative bias is an increasing function of $\gamma$, 
it goes from negative values at $\gamma=0$ to positive values for $\gamma>2$. This behavior can be
understood from the fact that for a given value of the reconstructed energy proton events come from 
larger values of the true energy but iron events come from smaller values, therefore, for a power law 
energy spectrum the iron events have a larger weight, which increases with $\gamma$ moving the mean 
value towards the one corresponding to iron nuclei. In this way, after increasing $\gamma$ sufficiently 
the relative bias becomes positive.  

For every $\gamma\neq2$, the relative bias has an extreme point placed at an intermediate value of the 
proton abundance. This extreme value is a maximum if $\gamma>2$ and it is a minimum if $\gamma<2$. The 
expression for the proton abundance corresponding to the extreme point can be obtained from 
Eq.~(\ref{RbSC}), which is given by,   
\begin{equation}
c_p^{ext}=\frac{R_\mu R_S^{\gamma-1}-\sqrt{R_\mu R_S^{\gamma-1}}}{R_\mu R_S^{\gamma-1}-1},
\end{equation}
which is valid for $R_S\neq1$, $R_\mu \neq R_S$, and $\gamma\neq2$. As can be seen from figure 
\ref{BiasSC}, $c_p^{ext}$ varies very slowly with $\gamma$, in fact it goes from $\sim 0.54$ at 
$\gamma=0$ to $\sim 0.58$ at $\gamma=3.3$.

\end{document}